# Cryptography: Against AI and QAI Odds

**Sheetal Harris, Hassan Jalil Hadi, Umer Zukaib**
Cyber Science and Engineering Cyberspace Security
Wuhan University, China

*Abstract*— **Artificial Intelligence (AI) presents prodigious technological prospects for development, however, all that glitters is not gold! The cyber-world faces the worst nightmare with the advent of AI and quantum computers. Together with Quantum Artificial Intelligence (QAI), they pose a catastrophic threat to modern cryptography. It would also increase the capability of cryptanalysts manifold, with its built-in persistent and extensive predictive intelligence. This prediction ability incapacitates the constrained message space in device cryptography. With the comparison of these assumptions and the intercepted ciphertext, the code-cracking process will considerably accelerate. Before the vigorous and robust developments in AI, we have never faced and never had to prepare for such a plaintext-originating attack. The supremacy of AI can be challenged by creating ciphertexts that would give the AI attacker erroneous responses stymied by randomness and misdirect them. AI threat is deterred by deviating from the conventional use of small, known-size keys and pattern-loaded ciphers. The strategy is vested in implementing larger secret size keys, supplemented by ad-hoc unilateral randomness of unbound limitations and a pattern-devoid technique. The very large key size can be handled with low processing and computational burden to achieve desired unicity distances. The strategy against AI odds is feasible by implementing non-algorithmic randomness, large and inexpensive memory chips, and wide-area communication networks. The strength of AI, i.e., randomness and pattern detection can be used to generate highly optimized ciphers and algorithms. These pattern-devoid, randomness-rich ciphers also provide a timely and plausible solution for NIST's proactive approach toward the quantum challenge. AI cryptanalysis mitigation tactics provide security for medical, AIoT devices, and the military since it avoids the computational load of traditional ciphers. Therefore, a corresponding cryptographic solution can encounter the looming cyber threat.**

*Keywords*—**Cryptography, Artificial Intelligence, Quantum Computing, Quantum Artificial Intelligence, AI Cryptanalysis, Cyber Security**

## 1. INTRODUCTION

The interconnected world extensively depends on secure Internet services for multifaceted purposes, such as email, social networking, online banking and e-commerce [6]. The https protocol employs Secure Socket Layer (SSL) at 128-bit encryption to safeguard web traffic for secure communication. In the realm of the cyber-world, AI is a threat to traditional cryptography, where the complexity, frequency and robustness of cyberattacks are inexhaustible. Kaspersky[1] Cybermap in Figure 1. illustrates live cyberattacks worldwide using ports. The ports serve numerous computing services, e.g., email and https; which are constantly attacked by hackers.

AI leverages complex algorithms, learning and problem-solving techniques, where its level of deduction has proved its competence beyond human cognitive abilities. The widespread applications of AI in medical systems [9], remote sensing [7], and cloud and edge computing [5] have broadened a new horizon for researchers. Over the years, traditional cryptography and AI have evolved in their mutual dichotomy [8]. However, the co-existence of AI and Cryptography may follow a two-pronged approach. AI can improve existing cryptographic schemes, their efficacy, security and confidentiality. Contrarily, AI can be used as a modern cryptanalytic tool.

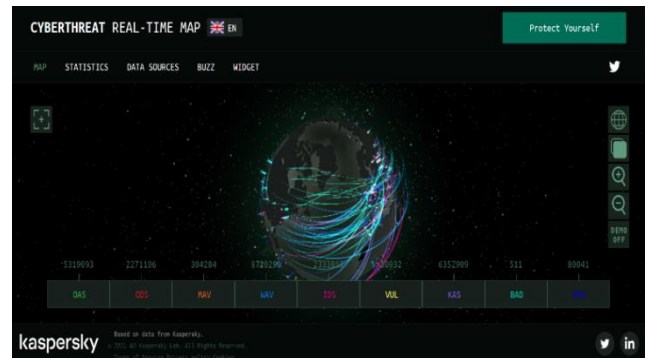

Figure 1. Kaspersky Global Live-Attack Map

The well-known cryptopunk, Julian Assange[2], validates the significance of cryptography as an army of a state. He claims that it protects the independence and objectivity of an organization and its Information Systems (IS). Cryptography techniques use algorithms, secret keys, mathematical problems, structures and intricate transformations to preserve data confidentiality during storage or transmission against illegitimate access [2]. The cryptographic techniques aim to protect communications in compliance with the CIA-triad (Confidentiality, Integrity, and Availability) and non-repudiation [1].

Cryptanalysis is the technique to identify vulnerabilities in system architecture, encryption algorithms and implementation that are exploited to break into cryptographic systems. It is the art to decrypt the encrypted messages without the encryption key by breaking the ciphers using mathematical analysis and algorithms [4]. Since cryptography safeguards communication and IS from impervious attacks, cryptanalysis techniques are employed to hack into the system for unauthorized access to communications [3]. To reinforce weak algorithms, researchers also use cryptanalysis techniques to identify design and flawed algorithms. Contrarily, the attackers use cryptanalysis to commit cybercrimes, whereas white-hat hackers use it to perform penetration testing to identify system vulnerabilities and security thresholds.

AI has emerged as the most powerful tool, for the AI sky is not the limit anymore. The Godfather of AI, Geoffrey Hinton, has alerted the world about the catastrophic effects and implications on the existence of human beings and IS[3]. With the development of QAI systems, the situation may worsen, and together they pose a great threat to traditional cryptography. Moreover, the fact cannot be denied that it will enhance the attacker's capability to launch impervious attacks on a large scale. The escalation of attacks is attributed to the fact that computers are faster and more capable. Quantum computers can execute exponentially more calculations than classical computers. Whereas, the computations in classic computers are limited to the number of cores in their processors. For instance, the supercomputer *"Sunway TaihuLight"*, with 10,649,600 processor cores can perform 21000 calculations, which can be performed by the quantum computer, D-Wave X2 [14] with 1000 qubits (one processor).

---

[1] https://cybermap.kaspersky.com/
[2] https://www.brainyquote.com/quotes/julian_assange_602821
[3] https://www.abc.net.au/news/2023-05-03/geoffrey-hinton-godfather-of-ai-quits-google-with-danger-warning/102297868

The security of cryptographic systems pivots on the complexity and randomness of computational algorithms. Claude Shannon's notion of cryptanalysis as *"an operation that changes the a-priori probability distribution over the message space to posteriori distribution of smaller entropy"* is challenged by AI [4]. Traditional cryptanalysts have been replaced with AI. It has valuable knowledge and lower entropy compared to Shannon's idea of no prior knowledge about plaintext and higher entropy.

The dynamic actions performed by AI result in acquiring valuable information even if limited information is available [12]. AI gathers knowledge from the available bits and pieces anticipated as partially relevant by the attackers. The power of AI has been underestimated by its developers. The astonishing fact remains that AI would be a useful tool for attackers in extracting hidden patterns to surpass the current cryptographic algorithms and protocols to break codes.

This pre-cryptographic stage involves the vulnerability of the circumstantial information shared by the transmitter. AI as a cryptanalytic tool potentially processes circumstantial information and exploits patterns related to a particular user or the transmitter. Once, the information is processed, it infers plaintext without the key. The study [11] validates that the ciphertext and its corresponding plaintext in a cipher are prone to AI cryptanalysis. Therefore, each ciphertext and its matching plaintext can be compromised. AI pursues the matching plaintext for a given ciphertext, which is easier than decoding the ciphertext and retrieving a key. Therefore, the challenging situation in the presence of AI and QAI demands effective proactive techniques and methods to strengthen current cryptographic defences.

The previous research [15, 16] in AI focuses on the development of AI. However, to this day, the development of cryptography in the presence of AI as a cryptanalytic tool is an under-researched area. Therefore, this study demonstrates the current state of modern cryptography. It also leverages measures to curb the threat of AI as cryptanalytic tools by using randomness and pattern-devoid cryptography.

## 2. MOTIVATION

AI has evolved as the process that stimulates the cognitive ability of machines as a human intelligence process. The emergence of AI as a cryptanalytic tool in the presence of QAI and quantum computers poses a threat to current cryptographic systems. The attackers can also utilize the power of AI to surpass traditional cryptography algorithms and launch impermeable attacks. Tampering with cryptographic algorithms will have a catastrophic impact on the cryptographic systems and PII. In unison with the NIST's proactive approach towards the quantum challenge, AI's ability to learn and evaluate must be challenged with a plausible solution. Therefore, in this paper, we will propose a proactive approach to restrict the capability of AI using its ability against it as a cyber-security defence for cryptographic systems.

## 3. RESEARCH CONTRIBUTION

AI cryptanalytic tool poses a threat to the existing cryptographic systems. The attackers can misuse the efficacy of AI for knowledge acquisition and pattern recognition. Cryptanalysts can widely harness the power of AI for their illegitimate purposes.

- We have identified challenges and threats AI poses to the current cryptographic systems and algorithms.
- The effectiveness of BitMap and BitFlip techniques for safe communication in cryptographic systems is demonstrated in the study.
- AI-assisted Cryptography using randomness and pattern devoid cryptography determines how AI can be utilized to strengthen the existing cryptographic systems.

## 4. LITERATURE REVIEW

The applications of AI in cryptography are contemporary compared to AI applications in security. There is a scope for amelioration that may propose how AI threats to cryptography can be overcome using state-of-the-art techniques. This study includes research works; focused on AI in cryptography and how AI can be used against AI attacks on current cryptographic systems.

Modern cryptography and AI have evolved together over the years. The study [10] by Ronald Rivest determined that "Machine learning and cryptanalysis can be viewed as Sister fields since they share many of the same notions and concerns". The author stated their mutual challenges *"This problem can also be described as the problem of learning an unknown function (that is, the decryption function) from examples of its input/output behaviour and prior knowledge about the class of possible functions".* He maintained that AI has the potential to be used as a cryptanalysis tool to compare and identify the matching plaintext and ciphertext. The study [11] also suggested that it is much easier for AI to assess the identical plaintext and its corresponding ciphertext.

Multi-layer Perceptron (MLP) neural network was used by [28] to map the Simplified Data Encryption Standard (S-DES) behaviour. The researchers [31] used CBC mode to extract valuable information from the ciphertext without prior knowledge. ECB mode showed better performance and resulted in the acquisition of more information. The same method ML-based was used by [32, 33] for cryptanalysis. The study [34] indicates that the DL-based method is used to match plaintext and its corresponding ciphertext.

The DL-based model shows that the key of a lightweight block cipher can be successfully retrieved [11]. The research work [15] focuses on the pattern recognition capability of AI and suggests the pattern-devoid strategy for ciphertext. They further demonstrate that cryptanalysts can use AI to surpass cryptographic algorithms. However, AI can be used to generate strong encryption methods and encrypted data. The research by [17] also claims that ML can be widely used for cryptographic systems to strengthen cryptographic keys and secure encrypted traffic classification. The researchers further emphasise that ML can be widely used to execute side-channel attacks. It is also maintained that ML techniques should be used in cryptography.

The research work [18] addressed the issue of data privacy when data is transmitted between multiple parties. The gradient descent method was used to protect vertically and horizontally partitioned data and enhanced comprehensiveness in a general state. The study [19] shows that several cryptography techniques are computationally expensive. The efficacy of lightweight technique through randomization for reliable communications is demonstrated in their work.

The researchers in the study [20] suggested that a neural key exchange protocol can be designed by synchronizing the learning process of the two neural networks. They highlighted that synchronization delay results in a compromised neural key. Their method used the output frequency for degree assessment. Secondly, the hash function was used to assess effective synchronization by comparing it with an established degree threshold. The authors in [21] proposed the MTFLA algorithm based on fuzzy logic for the IoT ecosystem. They demonstrated the effectiveness of the design by detecting spoofing attacks, which was proved by simulation results. The proposed method analyses the probability distributions of received power discovered for the regions created for mobile (moving) users.

The key feature enhancement by modifying the AES algorithm was proposed by [22]. They showed that by increasing the encryption rate (i.e., 1000 blocks per second), 128 AES algorithm, data privacy and security are enhanced manifold. In their research work [23], the authors validate the effectiveness of the CNN-based system for enhanced network security. The results establish the efficacy of their proposed algorithm for malicious data identification.

The use of AI-based techniques for cryptography is an under-researched area to this day. The use of Neural Networks (NN) [15, 24, 25] for secure encryption and decryption is discussed. ML can also be used for enhanced cryptography, encrypted traffic classification and the public key cryptosystems through Tree Parity Machines (TPMs) [27, 28]. The other research works have also demonstrated the vulnerability of simplified ciphers and S-DES [26, 30]. The role of AI and quantum computing in cryptography is suggested by [29].

## 5. STRATEGIES AGAINST AI ODDS

The emerging AI threat to cryptography can be refuted using a novel approach, i.e. divergence from conventional cryptography. The a-priori list contains limited plaintext possibilities for a ciphertext and is used to launch the AI attack.

The same limited list can be changed into a terminal list of plaintext possibilities. The terminal should be developed in such a way that each terminal list corresponds to a ciphertext. The terminal list of plaintexts should contain the plaintext as similar as possible to the original a-priori list. The action is performed to obfuscate the ciphertext from the reach of AI. Therefore, the code developer can change the terminal list to secure this cryptanalytic vulnerability.

The novel approach differs from conventional cryptography, where ciphertext is deployed to generate the corresponding plaintext. However, this approach exhibits resistance to AI. The restrained ciphers can be created at the cost of communication burden and larger keys.

### 5.1. DECOY TOLERANT CIPHER

The proposed approach is based on decoy-tolerant ciphers [35]. Decoy tolerant cipher is defined as *"a cipher which quickly, easily and unequivocally distinguishes between proper message bits and decoy bits. The former it decrypts the latter it discards"*. The idea of decoy ciphers is based on the notion coined by Ron Rivest [36] based on the technique, *"chaffing and winnowing of wheat"*. It explains that, unlike typical ciphers, decoy-tolerant cipher only decrypts the bona fide material instead of deciphering the whole information. Consequently, it releases the communication burden and restricts unauthorized access and backdoors for law-enforcement agencies.

Therefore, to create resistance against AI, the ciphertext contains two different kinds of bits. The useful bits for the recipient (wheat) and the other bits (chaff) are just created to puzzle the attacker. The efficient classification results and closeness to the actual ciphertext is the key to secure ciphertext, which creates an impression of decoy for the cryptanalysts. The sender transmits the bit flow containing the useful and useless bits. The legitimate recipient can differentiate between the wheat and chaff proposed by Ron Rivest. Chaff contains random information as close to ciphertext as possible. The attacker considers the whole bit flow useful. In the bit flow, the useless bits are configured such that these only decrypt the ciphertext using a cryptographic key into plaintext. In this scenario, it is difficult for AI to ascertain the key used to decipher the ciphertext from different keys that are directed towards the same ciphertext. Therefore, by creating an identical terminal list closer to the a-priori list, the ciphertext will be inaccessible to AI and cryptanalysts.

They proposed that decoy-tolerant ciphers are devoid of encryption, only decrypt significant information and disregard the rest. He maintains, *"Winnowing does not employ encryption, and so does not have a decryption key"* [36]. Therefore, it ensures foolproof confidentiality to the extent that it restricts the back door through decryption key acquisition. The proposed strategy uses winnowing and steganography instead of using encryption.

The winnowing-based technique links the message with *"Message Authenticating Code – MAC"* and decoy bits (chaff). The counterfeit bits are added to puzzle the AI cryptanalyst. The source and content of the transmitted message are shared along with a secret key. The comparison of *"MAC (such as HMAC-SHA1 – random function)"* authenticates the message otherwise it is discarded.

Therefore, it used the concept of a one-time pad. For a longer message, the receiver identifies the relevant message through the added serial number and disregards the repetitions or bogus packets. For example, the message with random bits, MAC and serial number will appear as:
(1,Hi Stella,341245)
(2,Hi John,236790)
(3,Are you coming,645859)
(4,Are you going,338457)
(5,to restaurant,457853)
(6,to the movie,346280).

This message contains a tailored a-priori list, invalid MACs and serial numbers that can be deciphered only with the secret key. The transmitted message can be identified using the secret key shared between the users, otherwise, it will be discarded if the authentication fails. As suggested by [36], *"The chances of creating a good packet are one in $2^{64}$—approximately one in $10^{19}$—which is effectively negligible"*. Therefore, to add more sophistication, the message is changed to bits, which will become as follows:
(1,0100,789654)
(2,0110,678956)
(3,0101,453426) and so on.

When this message is changed into 64 bits, it will add more security to it without being encrypted and thus, provides a safe haven from AI cryptanalysis attacks.

### 5.2. AN EXTENDED TERMINAL LIST

The core idea of an extended terminal list is to create a mimic list and guard against AI and QAI threats. Imitating the a-priori list as close to the original a-priori list will establish an element of confusion for AI. The combined bit flow of a-priori list and mimics will mitigate the threat of adversaries and illegitimate access.

Suppose the a-priori list contains n number of elements in a plaintext, A*1, A*2, A*3, .... A*n. The a-priori list will be changed into a cipher. Therefore, the terminal list be tailored in such a manner that the candidate in the a-priori list will be imitated. The more the terminal list resembles the a-priori list, the more secure it is from illegitimate access. Therefore, the attack on the ciphertext will be fruitless.

Therefore, using a unique bit string for each candidate, Pb, P2, P3, …. Pq, the decoy cipher will be generated. If the sender transmits a string abc, where c points to candidate b in the string. The decoy cipher will be the bit string dbc for the attackers. Therefore, except Pb rest is gibberish. Thus, the bit flow x can be transmitted. The legitimate user can decipher the x messages, where (x-1) will be ruses to misguide the cryptanalyst, which is AI in this scenario.

### 5.3. LARGE SIZE KEYS

Claude Shannon proposed that the ciphertext will be secure from cryptanalysis attacks if the terminal list is identical to the a-priori list provided the processed message and the key size is the same [13]. For a smaller key than the processed message, the corresponding terminal list will be asymmetrical than the a-priori list. Therefore, data privacy and confidentiality depend upon the key size that can be adjusted using an identical terminal list to the a-priori list.

For example, AI identifies seven high-probability candidates in the a-priori list. The plaintext candidates A1, A2, A3, …. A7 with the corresponding keys (K1, K2, K3, ….. K7) and probabilities are illustrated in Figure 2. These keys can be used to decipher the corresponding ciphertext of only one plaintext from AI identified list. Whereas, this bit flow contains the plaintext and corresponding bogus terminal list. Therefore, the element of surprise and confusion alleviate the threat of AI.

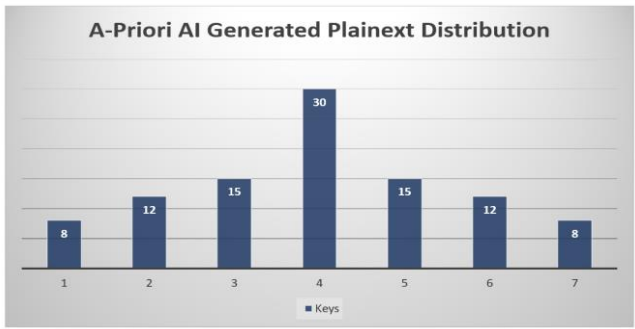

Figure 2: AI Generated Plaintext Distribution

The key size is further reduced and the corresponding probabilities are shown in Figure 3. Therefore, in Alpha cipher, plaintext candidates A2, A3, …. A6 in terminal list corresponds to keys K2, K3, …. K6. The terminal list will be more similar to the a-priori list of candidates.

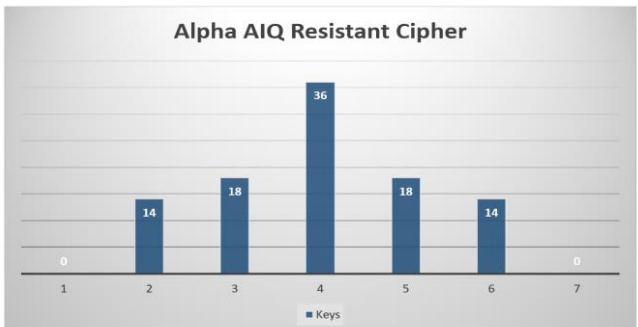

Figure 3: Alpha Cipher

AI will further reduce the cipher size by dropping the irrelevant keys as shown in Figure 4. This effort will be futile since the high probability entities are unsurpassed and AI cannot identify the corresponding plaintext from a weaker ciphertext. The bases of this unrivalled ciphertext are linked with the similarity between the a-priori list and the extended terminal list. Any discrepancy in this regard will pose an AI threat to cryptography.

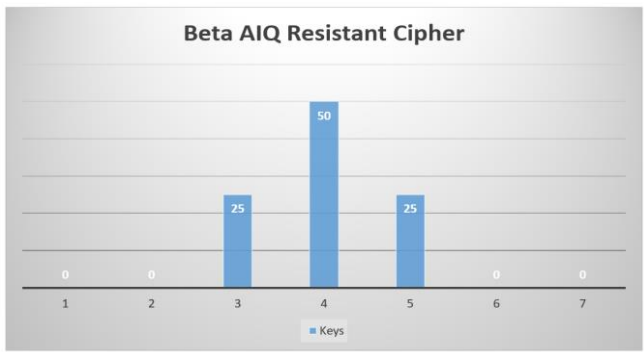

Figure 4: Beta Cipher

### 6. PATTERN DEVOID CRYPTOGRAPHY

The Godfather of AI, Geoffrey Hinton[4] has stressed the growing intelligence of AI that can surpass human intelligence. He mentioned that AI can identify hidden patterns and formulate intelligence and reasoning beyond human visibility and intelligence. Therefore, the existing pattern-loaded complex ciphers can be compromised by the cryptanalysts. The current computers and Turing machines lack randomness, which is the absence of patterns. Contrarily, emerging quantum computers perform various tasks using randomness. Hence, quantum computers pose a threat to the existing IS and cryptographic systems. Randomness can be used to alleviate AI and QAI threats.

The plaintext is pattern rich before encryption in existing cryptographic systems, where complex patterns are used to change it into a ciphertext. The cryptanalyst compares the ciphertext with plaintext and thus, intercepts the hidden message. According to Claude Shannon's theory, *"Cryptanalysts have no or limited knowledge about the a-priori list"* [13]. AI overrules this traditional approach. It identifies and analyses the pattern-rich a-priori list. Therefore, the security of the plaintext depends upon the concealment of the pattern that was used to convert the plaintext into a ciphertext [4].

To exceed the security threshold of classic ciphers against the threat of AI and QAI, sophisticated use of shared and unilateral randomness is proposed. Gilbert S. Vernam [38] proposed pattern-devoid cipher in 1917, i.e., also known as one-time pad cipher. The security of existing cryptographic systems can be increased using Trans-Vernam Cipher and modern technology. Vernam cipher is based on, *"the randomness of key and not mathematical properties that can be hacked"*. Thus, the cipher without patterns with random cryptographic keys and ad-hoc-based communication strategy can be used to protect the cryptographic system against AI and QAI threats.

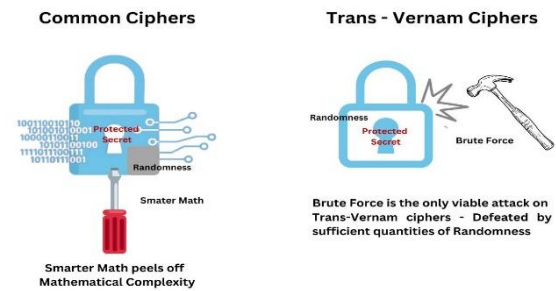

Figure 5: Traditional vs Trans-Vernam Ciphers

Trans-Vernam ciphers[5] provide assorted corresponding ciphers and terminated list of plaintext. The ciphertext with the matching cryptographic key can only be decrypted. Whereas, it keeps the cryptanalyst puzzled and confused in identifying the corresponding a-priori list and cryptographic key [39] as illustrated in Figure 5. Therefore, these ciphers are secure, decoy-tolerant, easy to implement, conceal plaintext size and provide customised capabilities for creating extended terminal lists, unlike traditional cryptographic ciphers.

### 7. DECOY TOLERANT CIPHERS

The decoy-tolerant ciphers used to mitigate the threat of AI and QAI as discussed as follows:

#### 7.1. BitFlip

This decoy cipher is based on Vernam cipher's notion of randomness, which is easy to implement with the current advancement of computational load and storage space. Trans-Vernam ciphers provide security and confidentiality, where a-priori list is combined with randomness to generate ciphertext unlike complex algorithms in traditional cryptography [40]. The author suggests [35] that the Trans-Vernam cipher entails the concept of equivocation [42] based on a large unicity distance. This new strategy envisions better approaches towards the challenges of traditional cryptography. It will also benefit to address the issues of quantum computing, AI and QAI [41].

---

[4] https://www.wired.com/story/geoffrey-hinton-ai-chatgpt-dangers/

[5] https://www.dcode.fr/vernam-cipher

BitFlip [40, 41] is a Trans-Vernam polyalphabetic cipher, defined as *"any cipher based on substitution, using several substitution alphabets"*. Therefore, in simple terms, the bits in a bit string relate to many bits and in return, many bits revert to one bit, i.e., based on *"One-to-Many-Many-to-One (O2M-M2O) relations"*.

In an alphabet, each letter is denoted as a bit of length (l) in the bit string (message) and the distance between the bits is represented by Hamming distance (h). The decoy string (tailored a-priori list) of bits (d) and bits (decoy) that can be deciphered as more than one letter is disregarded by the legitimate receiver. However, the use of equivocation and similarity between the a-priori list and random bits keep the cryptanalyst mystified. Therefore, BitFlip cipher adds an element of stymied randomness at will and decoy by design.

Suppose the transmitter sends a message m to the receiver, where a letter (s*) denotes a bit string (s). In a traditional cipher, a bit string (s*) is transmitted, and the receiver gets the letter (s). The fact is hidden that the letter (s*) corresponds to the letter (s) in the bit string. Cryptanalysts use various methods to map the relationship between the plaintext and ciphertext and access the plaintext, e.g., frequency analysis. Using a BitFlip approach, we will have another corresponding string closer to (s). Let (t) be another different bit string. Now, the cipher will contain both strings (s and t) as inputs. The relation (C) between these bits will be generated as ΨC. The transmitter will share the bit string (s*) by sending (t) and the relationship between these bits will be given as ΨC (s, t) = 1.

To represent the O2M-M2O relationship between the bits, (t) will correspond to many strings $t_1, t_2, …t_i$ for a relation ΨC (s, $t_i$) = 1 and (t) will also have string $s_1, s_2, …s_j$ for a relation ΨC ($s_j$, t) = 1.

To distract the cryptanalyst, the message (m), each letter (s) will be transmitted bit by bit. For (n) number of letters, therefore, we will have an $s_1, s_2, …s_n$ string of letters.

For a letter ($s_i$), the corresponding letters from the (t) string, will be transmitted, such that $t_{i1}, t_{i2}, .... t_{ij}$ for the relation ΨC ($s_i, t_{ij}$) = 1, where j=1, 2, ... The relation ΨC ($s_k, t_{ij}$) =0, for j=1, 2, … represents any other gibberish letter for ($s_k$) where k ≠ i.

The recipient will evaluate the string based on the concept of *"chaffing and winnowing of wheat"*. If the above-stated conditions are fulfilled, the bits will be assessed, i.e., the transmitted string $t_{ij}$ corresponds to the letter $s_i$ and the relation C is evaluated as "1", which will be deciphered. While the rest of the bits will be regarded as bogus.

Thus, the corresponding tailored list closer to the a-priori list will be created, such that, these retain the decoy feature for the illicit users. The relation C will between the bits (s) and (t) will be based on the Hamming distance. For the letters (n) in the string, the distance (h) between the strings will be corresponding to Ψ (s, t) = 1, and it will be true when h:h = D (s, t). Therefore, the letter $s_i$ will be transmitted from the string $t_{ij}$ such that $h_i$ = H ($s_i, t_{ij}$) and $h_i$ ≠ H ($s_k, t_{ij}$) for k= 1, 2, ...(i-1), (i+1), .... n.

For both string (s) and (t), the tailored list will be generated based on the bit string length and hamming distance (h). Therefore, the longer the a-priori list, the longer will be the tailored list, such that |s| = |t|, where the size of the terminal list can be adjusted by controlling the size of corresponding bit strings.

### 7.2. BitMap

BitMap cipher [43] provides randomness and security by design against AI cryptanalyst attacks. The larger-size decoy cipher can be created from the a-priori list. The bit size for each corresponding letter is increased to maintain the security of the ciphertext. Secondly, as many bits correspond to a distinct letter in the bit string, it cannot be deciphered. BitMap maps the bits using various paths and only the corresponding decryption key can decipher the exact relation between the bits and ciphertext [4].

The path (relation) between bits can be mapped using different ways to reach the destination. Therefore, it represents a list of various travel ways or destinations. The map with a full description of visited places and traversed roads can be used to discern the relationship between the bits and ciphertext. However, if the map is inaccessible, the ciphertext will direct towards different plaintext in the extended tailored list [35].

Suppose, there are various roads, s, t, and u lead to destinations, k, l, m, and n. The initial roads are mapped and shown in Figure 6.

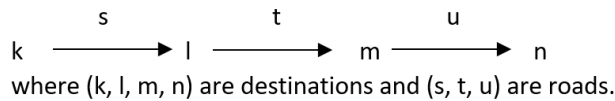

where (k, l, m, n) are destinations and (s, t, u) are roads.

Figure 6: Initial designated Roads and Destinations using BitMap

Similarly, various roads, g, h, and i also lead to destinations, k, l, m, and n, as shown in Figure 7.

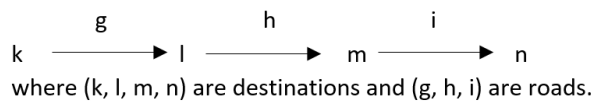

where (k, l, m, n) are destinations and (g, h, i) are roads.

Figure 7: Decoy Roads and Destinations using BitMap

The tailored terminal list will have various roads leading to the same destinations. The perfect decoy will be designed, provided the length of the priori-list is similar to that of the extended tailored list and the paths do not overlap. Since map size is unknown and the information about the departure and destination are anonymous. Therefore, AI and QAI cannot identify and map the relationship between the bits, it will be protected against the cryptanalyst attack. It can be compromised only if the travel paths intersect resulting in an impact on the low-probability bits.

BitMap is easy to implement. Base64 or ASCII are used to map the bit string (payload) that constitutes (n-1) letters. The a-prior list contains payload characters and all the letters are of the same size. The letters are joined together to form a string and in an a-priori list, these letters are repeated to form words. The idea is to add an n-th letter between each repeated letter so that the repetitions do not appear. Therefore, the tailored list will have no repetitions and all the letters are different. Whereas, by replacing the bits in the tailored list, the authorised recipient gains access to the deciphered message.

The same path names as illustrated in Figures 6 and 7 are also used to create a decoy for the illegitimate user using BitMap, which can be decrypted with the help of the map. However, AI and QAI cannot surpass the decoy tolerance generated by BitMap.

### 8. AI FOR CRYPTOGRAPHY

The strength of AI is one of the major threats to cryptography, i.e., pattern recognition in complex data. The patterns are used in contemporary cryptography for generating ciphertext, which can be changed into plaintext using corresponding keys. Therefore, the ciphertext represents large randomness that is created using little randomness that represents the key. AI cryptanalytic tool can be used to decipher these patterns. The approach pattern-devoid cryptography embodies the solution against AI threat [15]. The ciphertext of a corresponding plaintext must be generated as a complicated randomization using a one-way function. This indicates the possibility of AI applications for generating strong ciphers and encrypted data resistant against cryptanalysis attacks. Therefore, it is similar to outwitting your enemy and using the opponent's strength against itself.

Plaintext is converted into a ciphertext to generate a cipher in cryptography. For example, the network is fed with random numbers, trained and tuned using weights. In return, it will generate the corresponding cipher of the input random numbers. In the case of two communicating entities A and B, the input and selected weights represent the keys. The keys can be used for encryption and decryption and are exchanged using a key exchange protocol between entities A and B. Here, the network architecture and selected weights represent the encryption and decryption algorithms in the existing cryptography [28]. The method can be used to create the cipher from a small number of random elements as the input.

The technique is based on Evolutionary Computing widely used to provide highly optimized solutions for complex problems. It portrays the idea of biological evolution, i.e., the significant inherited traits evolve and forsake the insignificant characteristics. AI is trained using random numbers initially, and after each iteration, by using stochastic optimization [37], the desired solutions are used for the next iteration. This results in producing highly optimized solutions for the problems. Thus, highly optimized algorithms can be generated for cryptographic systems using this method.

## 9. CONCLUSION

The power of AI and QAI is infinite and inexhaustible. AI and QAI pose a catastrophic threat to existing cryptography because of their built-in robustness and pattern-detection capabilities. The developers of AI are mystified by its learning abilities to gain knowledge from available bits and pieces. Illegitimate users and attackers may use these capabilities to break into existing cryptographic systems. With the emergence of quantum computers, secure IS and communication have become a challenge. Thus, the existing cryptographic systems can easily be compromised using quantum computers. The randomness and pattern-devoid ciphers present the solution to mitigate the looming threat. In addition to that the extended tailored lists will keep AI cryptanalysts perplexed and confounded. The proposed decoy-tolerant ciphers are a step towards the NIST quantum challenge. Furthermore, by using the randomness and pattern detection capabilities of AI, highly optimized algorithms for cryptographic systems can be generated. Therefore, the proposed approach is to stay ahead and prepared and pursue a proactive approach rather than a reactive approach.